\documentclass[12pt]{article}
\pdfoutput=1
\usepackage{amsbsy}
\usepackage{amsfonts}
\usepackage{amsmath,amssymb}
\usepackage{bbold}
\usepackage{pdfpages}

\newcommand{\sect}[1]{\setcounter{equation}{0}\section{#1}}

\newcommand{\eq}{\begin{equation}}
\newcommand{\eqa}{\begin{eqnarray}}
\newcommand{\en}{\end{equation}}
\newcommand{\ena}{\end{eqnarray}}
\newcommand{\enn}{\nonumber \end{equation}}

\def\sk{\vskip .4cm}
\def\noi{\noindent}

\def\al{\alpha}
\def\be{\beta}

\let \part\partial

\def\epsi{\varepsilon}

\def\part{\partial}

\def\sk{\vskip .4cm}

\def\noi{\noindent}

\def\X0{X^0}

\def\al{\alpha}

\def\epsi{\varepsilon}

\def\L#1#2{ \La^{#1}_{~~~#2} }

\def\La{\Lambda}

\def\square{{\,\lower0.9pt\vbox{\hrule \hbox{\vrule height 0.2 cm
\hskip 0.2 cm \vrule height 0.2 cm}\hrule}\,}}

\def\Gtilde{\tilde G}

\def\Qbar{\overline{Q}}

\newcommand{\commas}{“}

\usepackage{amsfonts,slashed}
\usepackage{url}
\usepackage{latexsym}
\usepackage{amsmath}
\usepackage{amssymb}
\usepackage{indentfirst}
\usepackage{graphicx}
\usepackage{epsfig}
\usepackage{amsthm}
\usepackage{amsfonts}
\usepackage{indentfirst}
\usepackage{hyperref}
\usepackage{cite}
\usepackage{cancel}
\usepackage{tikz}
\usepackage{enumitem}

\usepackage{color}
\usepackage{slashed}

\usepackage{amsfonts,slashed}
\usepackage{url}
\usepackage{latexsym}
\usepackage{amsmath}
\usepackage{amssymb}
\usepackage{indentfirst}
\usepackage{graphicx}
\usepackage{epsfig}
\usepackage{amsthm}
\usepackage{amsfonts}
\usepackage{indentfirst}
\usepackage{cite}
\usepackage{cancel}
\usepackage{tikz}
\usepackage{enumitem}

\usepackage{color}
\usepackage{slashed}

\numberwithin{equation}{section}

\def\ii{{\rm i}}

\begin{document}

\begin{titlepage}
\vskip 2em
\begin{center}
{\Large \bf The $L_\infty$ structure of Free Differential Algebras and $d=11$ Supergravity} \\[3em]

\vskip 0.5cm

{\bf
L. Castellani${}^{1,2,4}$ and R. D' Auria${}^{3,4}$ }
\medskip

\vskip 0.5cm

{\sl ${}^{1}$Dipartimento di Scienze e Innovazione Tecnologica
\\Universit\`a del Piemonte Orientale, viale T. Michel 11, 15121 Alessandria, Italy\\ [.5em] ${}^{2}$INFN, Sezione di
Torino, via P. Giuria 1, 10125 Torino, Italy\\
[.5em] ${}^{3}$Politecnico di Torino, Corso Duca degli Abruzzi 24, 10129 Torino
\\[.5em]
${}^{4}$Regge Center for Algebra, Geometry and Theoretical Physics, via P. Giuria 1, 10125 Torino, Italy
\\[4em]}
\end{center}

\begin{abstract}
\sk

We discuss free differential algebras (FDA's), a generalization of the Cartan-Maurer equations for the group manifold vielbein, appropriate for theories containing $p$-forms ($p >1$). Their dual formulation is an extension of Lie algebras, called $L_\infty$ algebras, and we illustrate this duality in a simple example. Finally, we review the FDA structure and the dual $L_\infty$ structure of $d=11$ supergravity.

\end{abstract}

\vskip 2cm\sk\sk

\sk
 \noi \hrule \vskip .2cm \noi {\small
leonardo.castellani@uniupo.it, riccardo.dauria@polito.it}

\end{titlepage}

\newpage
\setcounter{page}{1}

\tableofcontents

\sect{Introduction}

The main idea of the group geometric (or group manifold) approach to the formulation of supergravity theories is to consider as basic fields  the {\sl components of the vielbein one-form} $\sigma^A=\sigma(z)^A_{~\Lambda} dz^\Lambda$ on the manifold of a Lie supergroup $G$,  {\small {\it A}} being an index in the $G$ Lie superalgebra, and $z^\Lambda$ the coordinates of the group manifold\footnote{The left-invariant vielbein one-form is defined as $g(z)^{-1} dg(z) = \sigma^A (z) T_A$, where $g(z)$ is the group element parametrized by $z$ and $T_A$ are the Lie algebra generators.}.
 This vielbein satisfies the Cartan--Maurer (CM) equations
 \eq
d \sigma^A + {1 \over 2} C^A_{BC} ~\sigma^B \wedge \sigma^C =0 \label{CM}
\en
where $C^A_{BC}$ are the structure constants of the $G$ Lie algebra. Tangent vectors
on $G$, dual to the vielbein $\sigma^A$, are denoted by $t_B$, so that $\sigma^A (t_B)=\delta^A_B$.
They satisfy the Lie algebra commutations
\eq
[t_A , t_B] = C^C_{AB} t_C
\en
The $G$ vielbein $\sigma^A (z)$  has a fixed dependence on the coordinates $z$, and hence cannot
be a dynamical field. We must then consider a ``soft" group manifold,
diffeomorphic to $G$ and denoted by ${\tilde G}$,
with a vielbein $\mu^A$ not satisfying anymore the CM equations. The amount of deformation from the original ``rigid" group manifold is measured by the {\sl curvature} two-form:
\eq
R^A \equiv d \mu^A + {1 \over 2} C^A_{BC} ~\mu^B \wedge \mu^C  \label{Gcurvature}
\en
Tangent vectors on $\Gtilde$, dual to the vielbein $\mu^A$, are still denoted by $t_B$ for simplicity of notation, and
satisfy $\mu^A (t_B)=\delta^A_B$.

Diffeomorphisms along tangent vectors $\epsi=\epsi^A t_A$ on $\Gtilde$ are generated by the Lie derivative $\ell_\epsi= \iota_\epsi d + d \iota_\epsi$, where $\iota_\epsi$ is the contraction operator.  When applied to the $\Gtilde$ vielbein, it generates its variation under diffeomorphisms:
\eq
\ell_\epsi \mu^A = d \epsi^A + C^A_{BC} \mu^B \epsi^C + \iota_\epsi R^A \label{Lieder}
\en
On the right-hand side one recognizes the $G$-covariant derivative of the infinitesimal parameter
 $\epsi^A$, plus a curvature term. When the curvature term vanishes, i.e. when
 $\iota_\epsi R^A=0$, the diffeomorphism
 takes the form of a {\it gauge transformation},  and the
 curvature is said to be {\it horizontal}
 along the $t_A$'s entering the sum in $\epsi=\epsi^A t_A$.
 Thus in group manifold geometry {\it gauge transformations} can be interpreted as {\it particular
 diffeomorphisms}, along those directions on which the curvatures are horizontal.

 This group geometric setting
 is particularly suited to supergravity theories, where local supersymmetry variations can be interpreted as diffeomorphisms in the super Poincar\'e group manifold, along the fermionic directions.

 Consider $G$ = superPoincar\'e group, and denote the vielbein on the $\Gtilde$ manifold as
$\mu^A = (V^a,\omega^{ab},\psi^\alpha)$. The index $A=(a,ab,\alpha)$ runs over the translations $P_a$, Lorentz rotations $M_{ab}$ and supersymmetry charges $\Qbar_\alpha$ of the superPoincar\'e Lie algebra (for details see Appendix A). The corresponding Cartan--Maurer equations read
 \eqa
& & R^{a}= dV^{a} - \omega^{a}_{~c} V^{c} - \frac{i }{2} \bar\psi\gamma^{a} \psi \equiv D V^a  - \frac{i }{2} \bar\psi\gamma^{a} \psi\label{RasuperPoincare}\\
& & R^{ab}=d \omega^{ab} - \omega^{a}_{~c} ~\omega^{cb} \label{RabsuperPoincare}\\
& & \rho= d \psi- \frac{1}{ 4} \omega^{ab} \gamma_{ab} \psi \equiv D \psi \label{rhosuperPoincare}
\ena
defining respectively the supertorsion, the
Lorentz curvature and the gravitino field strength. $D$ is the Lorentz covariant exterior derivative.
Wedge products between forms are understood when omitted.

 Supergravity actions in $d$ dimensions, invariant under local supersymmetry transformations, are obtained with an algorithmic procedure, as the integral of a $d$-form, ``living" on the whole supergroup (soft) manifold $\Gtilde$, but integrated on a $d$-dimensional {\it bosonic submanifold} of $\Gtilde$. This leads to an ordinary spacetime action containing the dynamical fields (and possibly also the auxiliary fields) of a $d$-dimensional supergravity theory.

The original references, where this approach was first proposed,
 are given in \cite{gm11}-\cite{gm14}. Reviews can be found
 in \cite{CDF} -\cite{handbook}, and \cite{EGH} is a standard reference for the use of differential forms in gravity and gauge theories.

 When supergravity multiplets contain $p$-forms (with $p>1$), Cartan--Maurer equations need
 to be generalized to {\sl free differential algebras} (FDA's).
 These were first introduced in the context of supergravity in reference \cite{DFd11}
 (see also \cite{DAuria:1982ada}) and were subsequently applied to various supergravity theories with higher order forms, see for example \cite{CDF}.

 Cartan-Maurer equations being dual to the Lie algebra commutation relations between the generators $T_A$, we may wonder what is the structure dual to the FDA equations.
 
 It turns out that this structure was introduced some ten years after reference \cite{DFd11} by mathematicians under the name of $L_\infty$ algebras, a generalization of graded Lie algebras with {\sl multibrackets} of any order \cite{Stasheff92,Lada:1992wc} (another way  of ``dualizing" FDA's was proposed in ref.s \cite{FDAdual1} --\cite{FDAdual4}, in terms of usual commutators and a generalized Lie derivative along antisymmetric tensors).
 
When mathematicians came in contact with the supergravity literature they realized that the FDA originally introduced in \cite{FDAdual1} had the same structure as $L_\infty$ algebras, albeit in a dual space formulation. The duality between FDA's and $L_\infty$ algebras was first discussed and elaborated  in a series of papers
\cite{Fiorenza:2013nha}, \cite{Fiorenza:2015gla}, \cite{Fiorenza:2016ypo}, \cite{Fiorenza:2019ckz}, \cite{nlab}, the last two references being very useful expository reviews.
These papers opened the way to a large number of interesting developments and observations \cite{Fiorenza:2016oki,Fiorenza:2017jqx,Fiorenza:2018ekd,Giotopoulos:2024jcr,Sati:2024klw}. \footnote{As an important example we recall the discussion of (higher T-) duality in string/M-theory as developed in references \cite{Fiorenza:2016oki,Fiorenza:2018ekd,Giotopoulos:2024jcr}.}

In the present paper, written mainly for physicists, we discuss further the duality between FDA's and $L_\infty$ algebras, and provide a simple pedagogical example where the duality is exposed in full detail. We then study this duality in the context of $d=11$ supergravity, where the FDA and $L_\infty$ dual structures are explicitly related.

 The paper is organized as follows. In Section 2 we recall how FDA's emerged in supergravity theories with higher forms. Section 3 contains a concise treatment of the FDA structure, and Section 4 deals with $L_\infty$ and its multibracket formulation. In Section 5 we prove the duality between FDA and
$L_\infty$, and verify it in the case of FDA1 (the simplest FDA, containing a 2-form). The cohomological construction of FDA's, starting from an ordinary Lie algebra, is recalled in Section 6. Finally, Sections 7 and 8 are devoted to the FDA and dual $L_\infty$ structures of $d=11$ supergravity. Appendix A contains the commutation relations of $d=4$ super-Poincar\'e Lie algebra, mostly to fix notations and normalizations.

\sect{The supergravity origin of FDA's }

When we consider supergravity theories in higher dimensions, for $6 \leq d\leq 11$, the  gravitational  supermultiplet contains in general antisymmetric tensors of rank greater than one, that is, $p$-forms with $p>1$ which cannot be part of the Lie algebra of a group G in the dual form of Cartan--Maurer equations.

 One has therefore to devise some way to incorporate higher $p$-forms in the framework of a new structure  generalizing a (graded) Lie algebra.

In the paper {\cite{DFd11} (see also \cite{CDF}) the authors find this way by defining a new structure which generalizes the Cartan--Maurer framework in such a way as to include also $p$-forms of any degree $p\geq 1$, besides the set of  1-forms spanning the Cartan--Maurer set. Inspired by the 
Cartan--Maurer structure of graded Lie algebras (GLA in the following), they consider a set of $p$-forms of various degrees and require that the \emph{d-exterior differential} of every $p$-form be expressed as a polynomial of (wedge) products of all possible $p$-forms in the set.  The integrability of this structure is obtained by the cohomological requirement $d^2=0 $, exactly as required for the Cartan--Maurer equations of an ordinary Lie Algebra, thus providing the generalization of the Jacobi identity.

 These structures were given the name  of \emph{Cartan integrable systems} (CIS). Later on, after the authors were made aware of a paper by Sullivan on {\sl free differential algebras} \cite{sullivan} (see also \cite{vanNieuwenhuizen:1982zf}), the name FDA was adopted in the supergravity literature.\footnote{Mathematicians also pointed out that the name of free differential algebras is not fully appropriate, since such structures are not \commas free" but only \emph{semi-free}, their underlying
 graded algebras being free. In the following we continue to call these structures FDA, the semifree character being understood.}

  Besides the importance of this FDA structure for the construction of $d=11$ supergravity in a  completely geometric framework, the introduction of FDA had an impact on the theory of (graded) Lie algebras.  Indeed the duals of FDA's were found to be an extension of graded Lie algebras to more general structures called $L_\infty$ algebras. The equivalence (duality) between FDA's and $L_\infty$ algebras will be made explicit in the next Section.

A further development in the FDA approach was obtained in the same reference  \cite{DFd11}. It was shown that the FDA of $d=11$ supergravity could be ``resolved"  in an ordinary Cartan--Maurer structure involving only 1-forms, at the cost of enlarging the original Lie algebra contained in the FDA. This allows in principle to trade the $L_\infty$ structure with an ordinary (but enlarged) graded Lie algebra. The possibility of a \emph{transmutation} of a $L_\infty$ algebra into an ordinary graded Lie algebra is an interesting result in the general theory of $L_\infty$ algebras which as far as we know has been discussed only recently in  \cite{Giotopoulos:2024ovz}. Recent references on the ``hidden" Lie algebra of $d=11$ supergravity can be found in \cite{Bandos:2004xw} - \cite{Andrianopoli:2024qwm}

  Finally, we mention that the FDA construction in the dual Grassmann space gives a natural extension to higher $p$-forms of the Chevalley--Eilenberg cohomology algebra, as shown in the following.

\sect {FDA structure}

In this Section, we describe the FDA construction in terms of the dual graded Grassman space $\mathfrak{g}{}^*$ {\cite{DFd11}. In the next Section we describe the structure of the $L_\infty$ algebra as defined in terms of multi-brackets. We then show explicitly that the two formulations are equivalent.

The use of the higher Chevalley--Eilenberg cohomology allows to construct a suitable FDA starting from the $d$-dimensional Poincar\'e algebra for any supergravity living in dimensions $6 \leq d\leq 11$. In Sections 7 and 8 we present this construction in the case of $d=11$. This choice is motivated by the fact that  supergravity in $d=11$ is not only the first theory where historically this new kind of higher tensor structures appeared, but also because $d=11$ supergravity is in a sense the most general supergravity theory. Indeed, from this theory, using dimensional reduction and T-duality, one can obtain all the $d\leq 10$ dimensional supergravity theories. Moreover $d=11$ can be considered as a low energy approximation of the so called M-theory.

Let us begin with the construction of the FDA.  We introduce on a manifold $\mathcal{M}_d$, whose dimension $d$ is not determined for the moment, a set of  $p$-forms $\{\Theta^{A(p)}\}$ of various degrees $1\leq p\leq p_{\mathrm{max}}$,
    where $A(p)$ is an index in a given representation of a structure group $G$, usually the Poincar\'e group in $d$ dimensions, \footnote{Actually the following construction can be also extended to the semisimple anti-de Sitter group by extending the Chevalley--Eilenberg cocycle to a cosmococycle, see reference \cite{gm13}.}
such that their exterior derivatives can be expressed as a polynomial in the set of
$\{\Theta^{A(p)}\}$ itself, with constant coefficients:
\begin{equation}
d{\Theta^{A(p)}}+\sum_{n=1}^{\infty}\,\frac{1}{n!} C^{A(p)}{}_{B_{1}(p_1)B_{2}(p_2)\dots B_n(p_n)} \Theta^{B_1(p_1)}\wedge\Theta^{B_2(p_2)} \wedge ... \wedge \Theta^{B_n(p_n)}\label{cis}
 \end{equation}
where $ C^{A(p)}{}_{B_{1}(p_1)B_{2}(p_2)\dots B_n(p_n)}$ are generalized structure constants with the same symmetry as induced by permuting the $\Theta$'s in the wedge product. Thus neighbour indices can be permuted with the rule:
\begin{equation}
B_i(p_i)\,B_{i+1}(p_{i+1}) = (-1)^{|B_i| |B_{i+1}|+ p_i p_{i+1}}B_{i+1}(p_{i+1})B_i(p_i)\,,\label{exchange}
\end{equation}
where $|B(p)|$ denotes the grading of the form $\Theta^{B(p)}$, i.e. 0 for a bosonic form,
1 for a fermionic form.

Moreover, since $d\Theta^{A(p)}$ is a $p+1$-form, the structure constants can be different from zero only if  $p_1+p_2+\dots +p_n= p+1$. We have extended the sum to infinity since for finite $p$ the sum truncates automatically to some integer $N$.

We now impose the \emph{integrability} of equation (\ref{cis}), namely $d^2=0$:
\begin{align}
 &d^2  {\Theta^{A(p)}}=-\sum_{n=1}^N \frac{1}{(n-1)!} \sum_{m=1}^N \frac{1}{m!}
C^{A(p)}{}_{B_{1(p_1)}B_{2(p_2)}\dots B_{n(p_n)}}\,C^{B_1(p_1)}_{D_{1(q_1)} D_{2(q_2)}\dots D_{m(q_m)}} \times\nonumber \\
& \Theta^{D_1(q_1)}\wedge \Theta^{D_2(q_2)}\wedge\dots \wedge\Theta^{D_m(q_m)}\wedge
\Theta^{B_2(p_2)}\wedge\dots \Theta^{B_n(p_n)}=0 \label{closure}.
\end{align}
The above closure condition is satisfied if the set of structure constants $C^{A(p)}{}_{B_{1}(p_1)B_{2}(p_2)\dots B_n(p_n)}$ satisfies the {\sl generalized Jacobi identities}:
\begin{align}
\sum_{n} \frac{1}{n!}\sum_{m}\frac{1}{m!} C^{A(p)}{}_{B_{1(p_1)}\bigl[B_{2(p_2)}\dots B_{n(p_n)}}\,C^{B_1(p_1)}{}_{D_{1(q_1)} D_{2(q_2)}\dots D_{m(q_m)}\bigr]}=0     \label{kos}
\end{align}
Here we denoted by $[...]$ the graded symmetrization of the indices, according to \eqref {exchange}.

We will return to this equation after describing the alternative construction first given in \cite{Stasheff92},\cite{Lada:1992wc} and \cite{nlab}.

\sect{$L_\infty$ structure}

  In this Section, we show that the FDA's have the same structure as the $L_\infty$ algebras, albeit in a dual language.

The FDA formalism lives in the Grassmann-algebra space of differential forms and can be considered as dual to the formalism based on the
    vector generators  $t_A$ of the GLA $\mathfrak{g}$.
    We illustrate this point with the
   example of an {\sl ordinary} Lie algebra, considering it from the two different perpectives.

   In one case we work in the \emph{antisymmetric space} $(V\otimes_a V)$ of Lie algebra generators $T_A$ and
  consider a set of two vectors  $(t_A,t_B)$,\footnote{ For the sake of simplicity we consider the basis of the generators $t_A$. By linearity it can be extended to any set of two vectors $(v_1,v_2) \equiv v_1 \otimes v_2 - v_2 \otimes v_1$.} living in  $(V\otimes_a V)$. We then introduce an operator $D$ called \emph{derivation}, with degree -1, mapping $(t_A,t_B)$ into a linear combination of the generators whose coefficients define a \emph{bracket}:
 \begin{align}
& D(t_A,t_B)  = -[t_A,t_B]^M\,t_M\equiv - C^M_{\;\;AB}\,t_M: \mathfrak{g} \otimes_a \mathfrak{g}\rightarrow \mathfrak{g}.
 \end{align}
On the other hand we may instead consider the Grassmann vector space $\mathfrak{g}^*$,  that is the \emph{dual co-algebra} whose generators are 1-forms $\sigma^A$, dual to the generators $t_A$ ($\sigma^A(t_B)=\delta^A_B $). In this case
the dual of the derivation, $D{}^*$, acts on the dual generators $\sigma ^A$ s as \emph{the exterior derivative operator} $d$:
 \begin{align}\label{MC1}
& d\sigma^A+\frac12 C^A_{BC} \,\sigma^B\wedge \sigma^C =0 :\mathfrak{g}{}^* \rightarrow \mathfrak{g}{}^*\wedge\mathfrak{g}{}^*.
 \end{align}
 The relation between these two dual languages is expressed by the relation:
 \begin{align}\label{defgen}
&d\sigma^A(t_B,t_C)=\sigma^A (D(t_B,t_C))=-\sigma^A\left([t_B,t_C]\right)\,.
 \end{align}
 Applying the exterior derivative to eq. (\ref{MC1}) and requiring $d^2=0$ yields the usual Jacobi identities $C^A_{B[C} C^C_{DE]}=0$. Can we derive them by requiring $D^2 =0$ on
 tangent vectors ? The answer is affirmative, provided we define the action of $D$ on
 $(V\otimes_a V \otimes_a V)$ as follows
 \eq
 D (v_1,v_2,v_3)=(D(v_1,v_2),v_3) + (D(v_2,v_3),v_1) +  (D(v_3,v_1),v_2)
 \en
 Then
 \eq
 D^2  (t_A,t_B,t_C) = - D (C^E_{AB} t_E, t_C)  =  C^E_{AB} C^F_{EC} t_F
 \en
 where antisymmetrization on the indices $A,B,C$ is understood. Therefore $D^2=0$ implies
 the Jacobi identities.

Following references \cite{Stasheff92}, \cite{Lada:1992wc}
\cite{{nlab}}, we can now generalize this example to more general algebras where \emph{higher} GLA structures appear.

Consider again the vector space $V$ of the graded Lie algebra $\mathfrak{g}$, and an element  of the \emph{n-fold graded antisymmetric} tensor product  $(V\otimes_a V\dots \otimes_a V)$, namely $(v_1,v_2,\dots,v_n)$.
In analogy to what has been discussed for the ordinary GLA, one then defines a map  given by the action of a ``differential"  $D$, a \emph{derivation} on elements $(v_1,v_2,\dots v_n)$. The grading of the vectors $v_i$ will be denoted by $|v_i|$ .\\
In this case, however, one has to generalize the derivation map $D$ as a sum of  ``differential operators"  $\ell_i$:
\begin{align}
  D=\ell_1+\ell_2 +\ell_3+\dots \label{Dsumell}
\end{align}
where each $\ell_i$ lowers $n$ by $i-1$,  i.e. the action of $\ell_i$ on an $n$-plet yields a
$(n-i+1)$-plet. 

An explicit example is treated in the next Section.

\subsection{The example of FDA1}

We can understand the reason for the decomposition (\ref{Dsumell})  by considering the simplest case of FDA, called
FDA1 in ref.s \cite{FDAdual1} -- \cite{FDAdual4}, where the Cartan--Maurer one-forms $\sigma^A$ of a group manifold G,
satisfying (\ref{MC1}), are supplemented by a single $p$-form $B$. Here for simplicity we take besides the 1-forms  $\sigma^A$ of the graded Lie algebra, a 2-form $B^i$ with a G-representation index $i$, satisfying
\eq
dB^i + C^i_{~Aj} \sigma^A \wedge B^j +  {1 \over 3!} C^i_{A_1 A_2 A_3} \sigma^{A_1} \wedge \sigma^{A_2} \wedge \sigma^{A_3}=0 \label{FDA1}
\en
Requiring $d^2 = 0$ on the FDA1 Maurer-Cartan identities (\ref{MC1}), (\ref{FDA1}) implies the FDA1 generalized Jacobi identities:
\eqa
& & C^A_{~C[B} C^C_{~DE]}=0  \label{Jac1}\\
& & C^i_{~Aj} C^j_{~Bk} - C^i_{~Bj} C^j_{~Ak} -C^C_{~AB} C^i_{~Ck} =0 \label{Jac2}\\
& & 2 C^j_{~[ABC} C^i_{~D]j} + 3 C^i_{E[AB} C^E_{~CD]} =0 \label{Jac3}
\ena
Consider now the dual formulation. A derivation $D$, acting on tangent vectors, is defined by the duality relations:
\eqa
& & d\sigma^A(t_B,t_C)=\sigma^A (D(t_B,t_C)) = - C^A_{\;\;BC}\\
& &dB^i (t_A,t_B,t_C)=B^i (D(t_A,t_B,t_C)) = - C^i_{~ABC}  \\
& &dB^i (t_A,t_j)=B^i (D(t_A,t_j)) = -C^i_{~Aj}
\ena
where in the last equality in every line we have used the FDA1 Maurer-Cartan equations, and $t_j$ is the bi-vector dual to the two-form $B^i$, i.e. $B^i (t_j)=\delta^i_j$.
We see then that the action of $D$ must be defined on couples and on triplets of
tangent vectors, and must incorporate the structure constants of the FDA1.  This justifies the decomposition of $D$ as a sum of ``differential operators“
$\ell_i$ :
\eq
D = \ell_1 + \ell_2 + \ell_3.
\en
Note that $D$ lowers the tensor degree by 1, being dual to $d$ that increases the form degree by 1.
Thus all $\ell_i$'s must lower the tensor degree by 1.

The properties of the $\ell_i$ are defined as follows:

1)  the action of $\ell_i$ on $i$-plets is determined by the FDA structure constants
\eqa
& & \ell_1 (t_A)=\ell_1 (t_i) =0 \\
& & \ell_2 (t_A, t_B) = - C^M_{\;\;AB}\,t_M \\
& & \ell_2 (t_A, t_i)= -C^j_{~Ai} t_j \\
& & \ell_3 (t_A,t_B,t_C)=  -  C^i_{~ABC} t_i
\ena

2) the action of $\ell_i$ on $n$-plets vanishes when $i > n$

3) the action of $\ell_i$ on $n$-plets when $i < n$ is given by
\eq
\ell_i (v_1,v_2,...,v_n)= \sum_\sigma \chi (\sigma,v) (\ell_i (v_{\sigma (1)},...,v_{\sigma (i)}),v_{\sigma (i+1)},...,v_{\sigma (n)})
\en
The {\sl graded signature} $\chi (\sigma,v)$ of the permutation $\sigma$ is defined by:
\eq
\chi (\sigma,v)= (-1)^{[\sigma ] + |v_j||v_k|}
\en
where $[\sigma $] is the usual signature of the permutation $\sigma$ and $|v_j||v_k|$ is due to  exchange of each pair of neighbour elements. Note that this phase is completely analogous to the phase generated by the exchange of two graded differential $p$-forms in equation  \eqref{exchange} .\\
It is then a straightforward exercise to verify that requiring $D^2=0$ on all possible
$n$-plets for any $n$ reproduces the FDA1 generalized Jacobi identities (\ref{Jac1})-(\ref{Jac3}).
For example,
\eqa
& & D^2 (t_A,t_B,t_C) = (\ell_1 + \ell_2 + \ell_3) (\ell_1 + \ell_2 + \ell_3) (t_A,t_B,t_C) =\nonumber  \\
& &
~~~~~~(\ell_1 + \ell_2 + \ell_3) ((\ell_2(t_A,t_B),t_C)+(\ell_2(t_B,t_C),t_A)+(\ell_2(t_C,t_A),t_B) + \ell_3
(t_A,t_B,t_C) )=\nonumber \\
& & ~~~~~~~ =\ell_2(\ell_2(t_A,t_B),t_C) +\ell_2(\ell_2(t_B,t_C),t_A )+\ell_2(\ell_2(t_C,t_A),t_B)= - 3 C^E_{~[AB} C^F_{~C]E} t_F \nonumber \\ \label{Dsquare}
\ena
while the second and third Jacobi identities (\ref{Jac2}), (\ref{Jac3})  are obtained by applying $D^2$ to the triplet $(t_A,t_B,t_i)$ and to the quartet $(t_A,t_B,t_C,t_D)$ respectively. To derive (\ref{Dsquare})
we have used the properties 1),2),3) of the $\ell_i$'s.
\sk
\noi {\bf Note 1:} the duals of the FDA's used in the supergravity constructions have a structure similar to the one of
FDA1, and contain also fermionic $p$-form fields (the fermionic character of a $p$-form
contributes to its overall grading). However these FDA's do not contain 0-forms, and therefore
supergravity theories with scalar fields have not been formulated completely in an FDA
framework. For this reason their $L_\infty$ duals always satisfy the requirement that $\ell_1$ vanishes on all tangent vectors $t_A, t_i$. Indeed only a 0-form $\phi$ can give rise to a $D$ operator acting only on one tangent vector $t$, because of the duality relation
\eq
d\phi (t_A) = \phi (D (t_A))
\en
where $D(t_A)$ is to be considered a ``$0$-vector", operating on 0-forms as a partial derivative.
One may consider generalizations of FDA's containing 0-forms: then in the dual $L_\infty$
the $\ell_1$ operators would be non-vanishing.
\sk
\noi {\bf Note 2:} the FDA1 algebra in the context of FDA - $\L_\infty$ duality has also been discussed
in ref. \cite{Reb}.

\subsection{The general case}

Let us deal now with the general case, i.e. the dual formulation of a generic FDA structure.
In the following we will denote the map
\begin{align}
\mathfrak{g}\otimes_a \dots \otimes_a\mathfrak{g}\rightarrow \mathfrak{g}
\end{align}
given by
$\ell_n\left( v_1, \dots, v_n \right) := [v_1, \dots, v_n]$ as a \emph{bracket.} Alternatively the bracket can be also identified with the \emph{ generalized structure constants}:
\begin{align}
 [t_{a_1}, \dots, t_{a_n}]^a t_a\equiv C^a_{~a_1,\dots,a_n} t_a.
\end{align}
Imposing the closure of the derivation map, namely $D^2=(l_1+l_2+l_3+...)^2=0$, yields
the generalized Jacobi identity. One obtains for any $n$:
\begin{equation}
\sum_{i+j=n +1}~\sum_{\sigma\,\in Sh(j,i-1)}\chi(\sigma,v_1,\dots v_n)\,(-1)^{j(i-1)}\,\ell_i\left(\ell_j\left(v_{\sigma (1)},v_{\sigma (2)}\dots v_{\sigma (j)}\right),v_{\sigma (j+1)},\dots v_{\sigma (n)}\right)=0 \label{sha}
\end{equation}
 that is also called \emph{strong homotopy Jacobi identity}, and characterizes $L_\infty$ algebras. The second sum in (\ref{sha}) is on the {\sl $(j,i-1)$ shuffle permutations}.
 A $(p,q)$ shuffle permutation, with $p+q=n$, is defined as a permutation $\sigma$ of
 $\{1,2,...n \}$ such that within the two subsets $\{\sigma(1),...,\sigma(p) \}$ and $\{\sigma(p+1),...,\sigma (n)\}$ one has $\sigma(i) < \sigma (i+1)$. Thus a shuffle of two ordered sets is a permutation of their ordered union which preserves the order of each of the given subsets.

\sect{The duality between FDA and $L_\infty$ Algebra}

We show here that the equations \eqref{kos} and \eqref{sha} actually coincide.
This equivalence has been first discussed in \cite{Lada1}, see also \cite{nlab} and 
\cite{Andrianopoli:2024qwm}.

We adopt the simplified formalism given in \cite{nlab}. Namely, we erase the indication of the form degree in the $\Theta^{B_i(p_i)}$ and rename $\Theta^A$ as $t^a$. Moreover we recall that the coefficients $C^{A(p)}{}_{B_{1}(p_1)B_{2}(p_2)\dots B_n(p_n)}$, which now take the form $C^a_{\;\;{b_1},{b_2}\dots,{b_n}}$, can be equivalently rewritten as
a bracket: \footnote{ Indeed, in analogy with equation \eqref{defgen}, the graded antisymmetric bracket $[t_{1},t_{2}\dots,t_n]$ can be identified with the generalized structure constant of the $L_\infty$ algebra.}
\begin{align}
  & C^a_{\;\;b_{1},b_{2}\dots,b_n} =[t_{b_1},t_{b_2},\dots,t_{b_n}]^a.
\end{align}
Therefore the generic FDA \eqref{cis} takes the simpler form:
\begin{equation}
 dt^a=-\sum_{k=1}^\infty \frac{1}{k!} [t_{a_1},\dots ,t_{a_k}]^a\, t^{a_1}\wedge\dots \wedge t^{a_k},
\end{equation}
 where the $t_a$ with lower indices are GLA generators  while $t^{a_1},\dots,t^{a_n}$ with upper indices are $p$-forms dual to GLA  generators. With these notations the $d^2$ operator on $t^a$ gives:
\begin{align}
&d\,d t^a= -d\sum_{k=1}^\infty \frac{1}{k!} [t_{a_1},\dots ,t_{a_k}]^a\, t^{a_1}\wedge\dots \wedge t^{a_k}= \nonumber \\
&^=\sum_{k,l}^\infty \frac{1}{(k-1)!\,l!} [[t_{b_1},\dots , t_{b_l}]\, ,t_{a_2},\dots ,t_{a_k}]^a\,t^{b_1}\wedge\dots\wedge t^{b_l}\wedge t^{a_2}\wedge\dots\wedge t^{a_k}=0\label{ddt}
\end{align}
which, given the previous identifications, coincides with \eqref{closure}.

One can at this point rewrite \eqref{ddt} (and consequently also \eqref{exchange}) in a slight different way so that it can be identified with the  \emph{strong homotopy algebra} condition \eqref{sha}. The key observation is that the wedge products of the $t^{a_i}$ (as for the equivalent   $\Theta^{B_i(p_i)}$-forms) project the nested brackets onto their graded antisymmetric components. This occurs because one can sum over all permutations $\sigma$ of the $k+l-1$ indices weighted with the Koszul phase $\chi (\sigma,t)$ of the permutation which was identified, in the FDA formalism, with the phase $(-1)^{B_i B_{i+1}+ {p_i p_{i+1}}}\,\equiv (-1)^{\left([\sigma]+|v_i||v_j|\right)}$ as shown in \eqref{exchange}.
It follows that we can rewrite the right-hand side of \eqref{ddt} as:
\begin{align}
& \sum_{k,l=1}^\infty \frac{1}{(k+l-1)!}\sum_{\sigma \in Sh(l,k-1)} \chi (\sigma,t) \frac{1}{(k-1)!\,l!}\,[[t_{b_1},\dots ,t_{b_l}],\,t_{a_2},\dots , t_{a_k}]^a \nonumber \\
& ~~~~~~~~~~~~~~~~~~~~~~~~~~~~~~~~~~~~~~~~~~~~~~~~~~~~~~~\cdot t^{b_1} \wedge\dots\wedge t^{b_l}\wedge t^{a_2}\wedge\dots t^{a_k}=0\,.\label{ddt0}
\end{align}
 The sum over all permutations can be decomposed into a sum over the $(l,k-1)$ \emph{shuffles}
and a sum over permutations inside the first $l$ and the last $k-1$ indices.
These latter permutations do not change the graded symmetry of the nested brackets, since the same permutation acts on the $t^{a_i}$ forms. As there are $(k-1)!\,l!$ of them, equation \eqref{ddt0}
can be rewritten as follows:
\begin{align}
& \sum_{k,l=1}^\infty \frac{1}{(k+l-1)!}\sum_{\sigma \in Sh \,(l,k-1)} \chi(\sigma, t) \,[[t_{a_1},\dots , t_{a_l}],\,t_{a_{l+1}},\dots ,
t_{a_{k+l-1}}]
t^{a_1}\wedge\dots\wedge t^{a_{k+l-1}}=0\,.\label{ddt2}
\end{align}
Therefore $d^2=0$ is equivalent to the conditions
\begin{align}
\sum_{k+l=n-1}\sum_{\sigma \in Sh\,(l,k-1)} \chi(\sigma,t) \,\Bigl[[t_{a_1},\dots, t_{a_l}], \,t_{a_{l+1}},\dots , t_{a_{k+l-1}}\Bigr] =0
\end{align}
that reproduce the strong homotopy identity \eqref{sha}, and therefore define an $L_{\infty}$ algebra.

In conclusion, FDA's, higher Chevalley--Eilenberg cohomology and $L_\infty$ algebra are all referring to the same algebraic structure.

\sect{The construction of the FDA}

We show in this section how to generate a graded FDA starting from an ordinary graded Lie algebra of a group $G$. The construction rests on the Chevalley--Eilenberg cohomology of a graded Lie algebra and more precisely on the relative cohomology of $H$-orthogonal cochains of the original GLA $\mathbb{g}$, $H$ being a subghroup of $G$.

Since our aim is to build a supergravity theory on a $d$-dimensional space-time, we will choose as starting point the Poincar\'e group in $d$ dimensions, which is a \emph{non-semisimple} group. On the other hand it is known that the original Chevalley--Eilenberg cohomology (CEC) only works in full generality for \emph{non-semisimple} groups, in the sense that for semisimple groups no non-trivial cocycle exists in degree one or two, while the first non trivial cocycle appears in degree three  in the trivial scalar representation.\footnote{  In fact the presence of the structure constants $C_{ABC}$ allow the non trivial cocycle $\Omega$:
\begin{align}
&\Omega=C_{ABC}\sigma^A \wedge \sigma^B\wedge \sigma^C.
\end{align}}

This fact seems to imply that the FDA construction cannot be applied to theories based on semisimple groups. This would exclude for example gravity or supergravity with a cosmological constant, based on semisimple groups as for example (anti)-de Sitter groups.

However in reference \cite{gm13} the authors have shown that it is possible to extend the CEC theory so as to include also semisimple algebras, obtaining in this way a generalization of the cocycle, called \emph{cosmococycle}, satisfying the Chevalley cohomology and also including semisimple groups on the same footing as nonsemisimple groups.

Therefore for theories with $
d < 11$, the same construction that we are going to present for the $d=11$  can be used using the generalized \emph{cosmococycle Chevalley cohomology}.

Here we limit ourselves to the description of $d=11$ supergravity where the natural space-time group is the Poincar\'e group since supersymmetry in $d=11$ forbids the presence of a cosmological term.

Let us describe how an FDA can be derived starting from the Maurer--Cartan equation of a non-semisimple graded Lie (co)algebra $\mathbb{g{}^*}$ of a given graded Lie group $G$, with subgroup $H\subset G$:
\begin{equation}
d\sigma^A +\frac12 C^A_{\;\;BC}\sigma^B\wedge \sigma^C=0.\label{MC}
\end{equation}
Consider an \emph{\emph{$H$-orthogonal}  relative Chevalley cochains complex} \footnote{An $H$-orthogonal cochain is such that it does not contain the $p$-form associated to $H$.  This is the situation when the Lie algebra has a subgroup $H$ which is a gauge symmetry of the theory. Thus a cochain $\Omega^i_{(n,p)}$ is $H$-orthogonal if $\iota_H \Omega^i_{(n,p)}=0=\iota_H \nabla^{n}\Omega^i_{(n,p)}=0$. For example if the theory we are constructing includes  Lorentz transformations $\rm SO(1,10)$
 (which is a subgroup of the (super)-Poincar\'e group in $d=11$), the gauge field $\omega^{ab}$ does not enter in the construction of the cochain.}.
Working with the \emph{relative} $H$-orthogonal Chevalley-Eilenberg algebra, we take as derivation operator the $H$-\emph{covariant derivative} $\mathcal{D}_{(H)}$, such that $(\mathcal{D}_{(H)})^2= R^{(H)A}T_A=0$, $R^{(H)A}$ being the $H$-curvature, which  vanishes in the FDA since the algebra physically represents the \emph{ground state} or \emph{vacuum} of the CE cohomology..

 A general element of the CE cohomology is a cochain that is a $p$-form polynomial of the type:
\begin{equation}\label{relCE}
\Omega^i_{(n,p)}=C^{i}_{{A_1},\dots,A_p}\,\sigma^{A_1}\wedge,\dots,\wedge \sigma^{A_p}\,,
\end{equation}
where $i=1,\cdots ,n$ runs in a $n$-dimensional  representation $D^{(n)}(T_A)^i{}_j$ of the Lie algebra generators $T_A$ of  $\mathbb{G}$,   ${A_1,\dots,A_p}$ being indices in the coadjoint representation and  $C^i_{A_1,\dots, A_p}$ constant invariant tensors of $G$.\\
  As we are starting from a GLA we use as $\sigma^{A_i}$ the 1-forms dual to the generators of the GLA.
 Next we introduce the $\mathbb{g}$-covariant derivative $\nabla^{(n)}$ acting on the $\Omega^i_{(n,p)}$:
\begin{equation}
(\nabla^{(n)})^i_j={d}\delta^i_j +\sigma^A\wedge D^{(n)}(T_A)^i_j\,.
\end{equation}\label{boundary}
Actually, in the case of  a Maurer-Cartan set of 1-forms, $\nabla^{(n)}$ coincides with the $\mathbb{g}$-covariant derivative computed at $R^A=0$.

Using equation \eqref{MC} we find:
\begin{equation}
\nabla^{(n)}\,\nabla^{(n)}=0
\end{equation}
and therefore $\nabla^{(n)}$ is a \emph{boundary operator}.

If the cochain is closed under $\nabla^{(n)}$ it is called a \emph{cocycle}, while a cochain is a \emph{coboundary} if there exists a cochain $\tilde{\Omega}^i_{(n,p-1)}$ such that
\begin{equation}
\Omega^i_{(n,p)} = \nabla^{(n)}\tilde {\Omega}^i_{(n,p-1)}.
\end{equation}
A cocycle which is not a coboundary is a representative of a \emph{Chevalley-Eilenberg cohomology class} of the Lie algebra.

Now, given a cocycle $\Omega^i_{(n,p)}$, we can introduce a new form $A^i_{(n,p-1)}$ and write the generalized Maurer-Cartan equation:
\begin{equation}
\nabla^{(n)}\,A^i_{(n,p-1)}+ \Omega^i_{(n,p)}=0.
\end{equation}\label{extend}
so that we may say that $A^i_{(n,p-1)}$ \emph{trivializes} the cocycle.

Adding this equation to \eqref{MC}, we obtain a \emph{higher Lie algebra}, actually a $L_p$ algebra (or a {\sl semifree graded differential algebra} of degree $p$}). Of course, the process can be iterated by considering a new set of cochains
 containing, besides the $\sigma^A$, also the $A^i_{(n,p-1)}$, namely:
\begin{align}
&\hat \Omega^{i}_{(n,p')}[\sigma,A]= C^i_{A_1,\dots,A_r\, i_1,\dots,i_s}
~\sigma^{A_1} \wedge \dots \wedge \sigma^{A_r}\wedge A^{i_1}_{(n_1,p_1)} \wedge \dots \wedge A^{i_s}_{(n_s,p_s)}.
\end{align}
If we can find new cocycles, say $\Omega^{\prime}$, in this enlarged cochain system, we then obtain correspondingly an enlarged FDA.

The process terminates when no new cocycles can be found, so that we have constructed the most general FDA  derived from the Lie algebra.

In the next subsection we apply this process to the construction of the FDA of the eleven dimensional supergravity, by starting from its underlying Lie algebra, namely the Lie algebra of the super-Poincar\'e group $\overline{{\rm Osp}(32|1))}$. \footnote{With the overline we mean the Inon\"u-Wigner contraction of $\mathrm{OSp}(32|1)$ to the super-Poincar\'e group. }

\sect{The FDA associated to the super-Poincar\'e algebra in {d=11}}\label{6.1}

The Maurer-Cartan equations of the D=11 super-Poincar\'e graded Lie algebra are given, in their dual form, in terms of
the set of 1-forms $\sigma^A=(\omega^{ab}, V^a,\psi^\alpha)$ (with $a,b,\dots=0,1,\cdots 10$, $\alpha=1,\cdots ,32$), where $\omega^{ab}$ is the $\mathrm{SO}(1,10)$ spin connection  and $E^{\hat a}=(V^a,\psi^\alpha)$ the supervielbein of $D=11$ superspace $\mathcal{M}_{11|32}$, $\psi$  being a  Majorana spinor in the 32-dimensional representation  of ${\rm Spin}(32)$. They read:
\begin{align}
 & d\,\omega^{ab}-\omega^a_{~c}\wedge \omega^{cb}=0, \label{poinc1}
\\
 &\mathcal{D}V^a -\frac{i}{2}\bar\psi\,\Gamma^a\wedge \psi=0\,.\label{poinc2}
\\
 &\mathcal{D}\psi\equiv d\psi-\frac14 \Gamma_{ab} \,\omega^{ab}\wedge \psi=0 \label{poinc3}\,.
\end{align}

In \eqref{poinc2}, $\mathcal{D}V^a= dV^a-\omega^{ab} \wedge V_b$  and $\mathcal{D}\psi$ denote the Lorentz covariant derivative of the bosonic and fermionic vielbein respectively.
Because the cohomology is $H$-orthogonal with respect to $H= \mathrm{SO}(1,10)$, the Chevalley cochains can be constructed using only  the supervielbein $V^a,\psi$.

Let us consider the trivial representation $D^{(0)}$, such that ${\nabla^{(0)}}$
reduces to the exterior derivative $d$. Constructing the CEC one finds that there is a non-trival cocycle of order four, namely:

 \begin{equation}
\Omega_{(V,\psi)}=\frac12 \bar\psi\wedge \Gamma^{ab}\psi\wedge V^a\wedge V^b.
\end{equation}
Indeed
\begin{align}
& d\Omega = \frac{i}{2}\bar\psi\wedge\Gamma^{ab}\psi\,\wedge \bar\psi \wedge\Gamma_a\psi\wedge V^b =0
\end{align}
where we have used equations \eqref{poinc2}, \eqref{poinc3} and the Fierz identity:
\begin{equation}\label{fierz11a}
\bar\psi\wedge \Gamma_{ab} \psi\wedge \bar\psi \wedge \Gamma^a\psi=0,
\end{equation}
which was proven in \cite{DFd11}. \footnote{The Fierz identity \eqref{fierz11a} expresses the fact that, in the symmetric product of four Spin(32) representations, the $\mathrm{SO}(1,10)$-vector representation is absent.}

According to the procedure previously explained, we can introduce a 3-form $A^{(3)}$ which locally \commas trivializes" the cocycle, writing:

 \begin{equation}
    dA^{(3)}-\frac12 \bar\psi \wedge \Gamma^{ab} \psi \wedge V^a \wedge V^b=0\,.\label{da3}
  \end{equation}
This equation, added to the Maurer-Cartan equations \eqref{poinc1},\eqref{poinc2} and \eqref{poinc3},
gives a FDA algebra, suitable for a geometrical construction of eleven-dimensional supergravity.
Indeed the 3-form $A^{(3)}$ provides exactly the degrees of freedom necessary to match bosonic and fermionic degrees of freedom of $d=11$ supergravity\footnote{Indeed in 11-dimensional space-time, the vielbein has on-shell
$\frac 12 d(d-3)=44 $   d.o.f., while the gravitino field has $2^{[d/2-1]}(d-3)= 128$ on-shell d.o.f.. Thus we need 84 more bosonic d.o.f. in order for the bosonic and fermionic d.o.f. to match. These are provided by an on-shell propagating 3-form  potential. Indeed, for a propagating antisymmetric tensor gauge potential of rank three, $A_{\mu\nu\rho}$, we have
$\frac{1}{3!} (d-2)(d-3)(d-4)=84$ d.o.f., so that the requirement is satisfied.}.

Now, after including $A^{(3)}$ in the enlarged set of CM forms, we can iterate the procedure in order to look for other non trivial cocycles.
We find that there is another cohomology class of order seven given by
 \begin{equation}
\Omega^\prime (V,\psi,A)=\frac{i}{2} \,\bar\psi \Gamma^{a_1,\dots,a_5}\wedge \psi \wedge V^{a_1}\wedge\dots\wedge V^{a_5}+\frac{15}{2}\bar\psi \wedge \Gamma^{ab}\psi \wedge V^a\wedge  V^b \wedge A^{(3)}.
\end{equation}
Indeed the differential of this expression is easily seen to vanish by use of  Fierz identities (see {\cite{CDF}, {\cite{DFd11}).

This allows to introduce a 6-form  $B^{(6)}$ locally \commas trivializing" the new cocycle $\Omega^\prime$:
\begin{align}
& d B^{(6)}=\frac{i}{2} \bar\psi\, \Gamma^{a_1\dots a_5}\wedge \psi \wedge V^{a_1}\dots \wedge V^{a_5}
 +\frac{15}{2} \bar\psi \wedge \Gamma^{ab}\,\psi \wedge V^a \wedge V^b \wedge A^{(3)}.\label{db6}
\end{align}
It can be verified that no new non trivial cocycles can be found.\\
\emph{Therefore we have found the most general FDA in superspace associated to the eleven dimensional super-Poincar\'e Lie Algebra}, whose generators are the Maurer--Cartan 1-forms $\sigma^A=(\omega^{ab},V^a, \psi)$, together with the 3-form $A^{(3)}$ and the 6-form $B^{(6)}$}. In dual terms, we have constructed an example of $L_n$ algebra, with the various $\ell_i$ given by
\eqa
& & \ell_1 (t_A)=\ell_1 (t(A)) =\ell_1 (t(B))=0 \\
& & \ell_2 (t_A,t_B) = -C^C_{AB} t_C \\
& & {1 \over 4!} l_4 (\Qbar_\alpha , \Qbar_\beta,P_a,P_b)={1 \over 2} (\Gamma_{ab})_{\alpha \beta} ~t(A) \\
& & {1 \over 5!} l_5 (\Qbar_\alpha , \Qbar_\beta ,P_a,P_b,t(A)) = {15 \over 2} (\Gamma_{ab})_{\alpha \beta}~ t(B)\\
& & {1 \over 7!} l_7 (\Qbar_\alpha , \Qbar_\beta ,P_{a_1},...P_{a_5}) = {i \over 2} (\Gamma_{a_1 ...a_5})_{\alpha \beta} ~t(B)
\ena
where $t_A$ and $C^C_{AB}$ are respectively the superPoincar\'e generators and structure constants
given in Appendix A, and $t(A)$, $t(B)$ are respectively the 3-vector and 6-vector dual to the
3-form $A$ and 6-form $B$.

\sect{Group-geometrical construction of $d=11$ supergravity}\label{6.2}

\par
Till now we have focused our attention to the basic properties of the geometry underlying the supergravities theories in $d \leq 11$ space-time dimensions. The result is a geometric construction of these theories, given in terms of
FDA's ( or $L_n$ algebras in the dual picture) corresponding to the properties of the ground state (or vacuum) of the theories, i.e. their "kinematic" structure.

 However, physical applications of the FDA require the introduction of \emph{curvatures}, that is field-strengths of the $p$-forms that enter the definition of the FDA. The standard procedure to arrive at physically propagating fields is well-known: one simply deforms the fields of the FDA in such a way that the right hand side of \eqref{poinc1}, \eqref{poinc2}, \eqref{poinc3}
 and \eqref{da3}, \eqref{db6} do not vanish any more, but {\sl define} the associated curvatures. In other words the left invariant fields of the FDA are no more left-invariant. \footnote{This construction corresponds to the generalization of the left-invariant 1-forms $\sigma^A$ of the Maurer--Cartan equations to the non left-invariant ``soft forms" $\mu^A$.}

 Let us denote by $\Pi^{A(p)}$ this deformed set of fields:
 \begin{align}
    \Pi^{A(p)}= \left(\omega^{ab},V^a,\psi,A^{(3),}
    B^{(6)}\right)\,.
\end{align}

 To build up the dynamical theory, all the concepts advocated for the geometrical construction of supergravity actions based on Maurer-Cartan equations can be straightforwardly extended to theories based on FDA's. One first introduces the (super)-curvatures $R^{A(p+1)}$ of the  $p$-forms $\Pi^{A (p)}$, corresponding to the deviation from zero of equations \eqref{cis} when the set of $ \Theta^{A (p)} $ is replaced by the \commas soft" forms $ \Pi^{A (p)} $. Therefore instead of equation \eqref{cis} we have:
\begin{equation}
R^{A(p+1)}\equiv d\Pi^{A(p)}+\sum_{i=1}^N \,\frac{1}{n!} C^{A(p)}{}_{B_{1(p_1)}B_{2(p_2)}\dots B_{n(p_n)}}
\Pi^{B_1(p_1)}\wedge\Pi^{B_2(p_2)}\wedge\dots \wedge\Pi^{B_n(p_n)}. \label {detetax}
 \end{equation}

Accordingly, the complete set of differential equations defining the curvatures of the FDA of $D=11$ supergravity is given by
\begin{align}
  \label{defcur1} R^a_{\;\;b} ~{\equiv} &~d\omega^a_{\,\,\,b}- \omega^a_{\,\,\,c} \wedge\omega^c_{\,\,\,b} {,} \\
 \label{defcur2} T^a ~ {\equiv}&~ \mathcal{D}V^a -\frac{{\ii}}{2} \overline \psi \Gamma^a \wedge \psi {,} \\
 \label{defcur3}  \rho~ {\equiv}&~ \mathcal D \psi {= d \psi - \frac{1}{4} \omega^{ab}}\wedge \Gamma_{ab} \psi \, \\
  \label{f4} F^{(4)} ~\equiv &~dA^{(3)} -\frac12\,\overline\psi\Gamma^{ab}\wedge \psi\wedge V^a\wedge V^b \\
 F^{(7)} ~\equiv &~ dB^{(6)}-\frac{i}{2} \overline\psi \Gamma^{a_1\dots a_5} \wedge \psi \wedge V^{a_1},\dots,\wedge V^{a_5}
 -\frac{15}{2} \overline\psi \wedge \Gamma^{ab}\psi \wedge V^a \wedge V^b \wedge A^{(3)} \nonumber\\
 &- 15 F^{(4)}\wedge A^{(3)}.
   \label{f7}  \end{align}
   The last term in \eqref{f7} (which is obviously zero in the vacuum) has been added to the right-hand side of eq. \eqref{f7} in order to have gauge invariance of the curvatures under the higher-form transformations:
   \begin{align}
   A^{(3)}&\rightarrow A^{(3)}+d\phi^{(2)}   \label{gaugea} \\
   B^{(6)}&\rightarrow B^{(6)}+d\lambda^{(5)}
   \label{gaugeb} \,,
   \end{align}
   where $\phi^{(2)}, \lambda^{(5)}$ are general 2-forms and 5-forms respectively.
Applying the exterior derivative to these equations yields the generalized Bianchi identities.

Given the definitions above, one then implements all requirements previously discussed for the construction of a geometric theory. In particular:
\begin{itemize}
\item The action functional is given in terms of a 11-form Lagrangian,   integrated over an eleven dimensional bosonic submanifold $\mathcal{M}_{11}$, immersed in the full superspace $\mathcal{M}_{11|32}$ parametrized by 11 bosonic and 32 fermionic coordinates, $(x^\mu;\theta^\alpha)$ respectively.
\item
The Lagrangian is completely \emph{geometric}: it is constructed in terms of $p$-forms and wedge products only, without the use of the Hodge-duality operator. As we have seen this implies that, even if  the Lagrangian is integrated on a submanifold of superspace (space-time), its being \emph{geometric} gives equations of motion valid on the full superspace.

\item We also add some symmetry conditions, namely the Lagrangian must be \emph{gauge invariant} under the gauge symmetries of the theory, which include the Lorentz $\mathrm{SO}(1,10)$ gauge symmetry together with the higher-form gauge invariances, eq.s \eqref{gaugea}, \eqref{gaugeb} . We add the obvious requirement that all terms scale and have the same parity properties as the Einstein-Cartan term.
\end{itemize}

\par We notice that the presence of the 6-form $B^{(6)}$, and of the associated curvature $F^{(7)}$ in the FDA, seems to violate the matching between bosonic and fermionic on-shell propagating d.o.f. However,
once the supersymmetric  and  gauge invariant Lagrangian has been written down, one finds that all the terms involving the 6-form  $B^{(6)}$ sum up to a total differential and therefore the field $B^{(6)}$ is not propagating. Furthermore, from the analysis of the Bianchi \commas identities" in superspace, it also follows that the components along the bosonic vielbein of the two field strengths $F^{(7)}_{a_1,\dots, a_7}$ and $F^{(4)}_{a_1,\dots, a_4}$ are actually Hodge dual to each other and therefore dynamically the degrees of freedom of $F^{(7)}_{a_1,\dots, a_7}$ are not independent from the ones of $F^{(4)}_{a_1,\dots, a_4}$. Physically, this means that once projected on space-time through $V^a_\mu$, the 7-form field strength $F^{(7)}_{\mu_1,\dots, \mu_7}$, is the \commas magnetic" Hodge dual of the \commas electric" field strength $F^{(4)}_{\mu_1,\dots, \mu_4}$.

 We do not report in this paper the explicit construction of the $d=11$ Lagrangian and/or the associated rheonomic parametrizations of the graded curvatures, satisfying on-shell the Bianchi identities in superspace. The detailed FDA-geometric derivation can be found in references \cite{DFd11} and \cite{CDF} (Vol 2, pag 861). We just report the resulting action :
\begin{align}
& \mathcal{A}=\int_{\mathcal{M}^{11|32}} \mathcal{L}
\end{align}
where $\mathcal{L}$ is a 11-form integrated on a 11-dimensional (bosonic) hypersurface immersed in the $\mathcal{M}^{11|32}$ superspace. The Lagrangian is \cite{DFd11,CDF}:\
\begin{align}
& \mathcal{L}=-\frac19 R^{a_1 a_2} V^{a_3} \wedge \dots \wedge V^{a_{11}} \epsilon_{a_1...a_{11}} +\frac{7}{30} T^a\wedge V_a \wedge \bar\psi \Gamma^{b_1 \dots b_5}\wedge\psi\,V^{b_6 \dots b_{11}}\epsilon_{b_1 \dots b_{11}} \nonumber \\ \nonumber
&+2\bar\rho\Gamma_{c_1 \dots c_8}\psi\wedge V^{c_1 \dots c_8}-84
F^{(4)}\wedge \left(\ii\bar\psi\Gamma_{b_1 \dots b_5}\wedge\psi\,V^{b_1 \dots b_{5}}- 10A^{(3)}\wedge \bar\psi \gamma_{ab} \psi V^{ab}\right)\\ \nonumber
& +\frac14 \bar\psi\Gamma^{a_1 a_2}\psi\wedge \bar\psi\Gamma^{a_3 a_4}\psi\wedge V^{a_5 \dots a_{11}} \epsilon_{a_1 \dots a_{11}}-210\, \bar\psi\Gamma^{a_1 a_2}\psi\wedge \bar\psi\Gamma^{a_3 a_4}\psi\wedge V^{a_1 \dots a_{4}} \wedge A^{(3)} \\
& -840\,F^{(4)}\wedge F^{(4)}\wedge A^{(3)} -\frac{1}{330} F_{a_1 \dots a_4}\,F^{a_1 \dots  a_4}V^{c_1\dots c_{11}}\epsilon_{c_1\dots c_{11}} +2 F_{a_1 \dots  a_4}\,F^{(4)} V^{a_5\dots a_{11}}\epsilon_{a_1 \dots a_{11}}.
\end{align}\label{supcur}
This is the Lagrangian on superspace where the fields still depend on $\theta$ and $d\theta$. Varying the fields we obtain two kinds of equations:

\noi 1) equations containing only  bosonic supervielbeins $V \wedge \dots V$, giving the equations of motion of the curvature components $\rho_{ab}, F_{a_1\dots a_4},R^{ab}_{~~cd} $:
 \begin{align}
   &\Gamma^{abc}\,\rho _{bc}=0\\
   &\mathcal{D}_m\, F^{m c_1 c_2 c_3}-\frac{1}{96}F_{a_1\dots a_4}\,F_{a_5\dots a_8}\epsilon^{a_1\dots a_8 c_1 c_2 c_3}=0\\
   & R^{am}_{~~bm}-\frac12 \delta^a_b R^{mn}_{~~mn}-3 F^{a c_1c_2c_3}\,F_{b c_1c_2c_3}
   - {3 \over 8} \delta^a_b F_{c_1 c_2 c_3 c_4} F^{c_1 c_2 c_3 c_4}=0 \label{racontract}.
 \end{align}
 2) equations of motion containing at least one fermionic vielbein $\psi$,  giving the on-shell parametrization of the curvatures:
 \begin{align}
  & T^a=0 \label{ti}\\
   & F^{(4)}=  F_{a{_1} \dots a_{4}} V^{a_1\dots a_4} \label{effe}\\
   & \rho=\rho_{ab} V^a\wedge V^b\,+ \frac{\ii}{3}\left(\Gamma^{b_1b_2b_3} F_{ab_1b_2b_3}-\frac18 \Gamma^{a b_1\dots b_4 } F_{b_{1}\dots b_{4}}\right) \psi\wedge V^a \label{ro},\\
   & R^{ab}=R^{ab}_{~~mn}\,V^m\wedge V^n+\bar \Theta^{ab}_{\;c}\psi\,V^c+
   \bar\psi\Gamma_{mn}\psi\,F^{mnab} +\frac{1}{4!}\bar\psi\Gamma^{ab a_1\dots a_4}\psi\,F_{a_1\dots a_4}\label{ra},
\end{align}
with $\Theta^{ab}_{\;c}=2\ii\Gamma_{[a}\,\rho_{b]c} -\ii\Gamma_c \rho_{ab}$.

In order to obtain the equations of motion on the eleven dimensional space-time we must consider that the components of the curvatures along the superspace vielbeins cannot be reduced to the space-time components just by using the relation $V^a= V^a_{\mu}\,dx^\mu$,  since the superspace vielbein has further components along the Grassmann coordinates
 $ d\theta^\alpha$. The right projection on space-time coordinates can be obtained as follows. Consider any curvature $R^A= (R^{ab},T^a, F^{(4)},\rho)$ and perform its projection along the coordinate differentials using its on-shell  parametrization as given in the previous equations. Taking as an example the curvature $\rho$ one obtains:
\begin{align}
&\rho_{\mu\nu}=\rho_{ab}V^a_\mu\,V^b_\nu - \frac{\ii}{3}\Gamma^{b_1 b_2b_3} F_{[ b_1 b_2 b_3 [\mu}\psi_{\nu]}- \ii\frac{1}{4!} \Gamma_{ b_1\dots b_4 [\mu} F^{b_1\dots b_4}\psi_{\nu]}.
\end{align}
where the term $\hat \rho_{\mu\nu}= {\rho}_{ab}V^a_\mu\,V^b_\nu$ defines the \emph{supercovariant field-strength} of $\rho$. As a consequence the gravitino equation of motion is:
\eq
     \Gamma^{a\mu\nu} \hat \rho_{\mu\nu} =  \Gamma^{a\mu\nu}  \left( \rho_{\mu\nu}+
     \frac{\ii}{3}\Gamma^{b_1 b_2 b_3} F_{b_1 b_2 b_3[\mu}\psi_{\nu]} +\ii\frac{1}{4!}
      \Gamma_{ b_1\dots b_4 [\mu} F^{b_1\dots b_4}\psi_{\nu]}\right) = 0.
 \en
Note that the $F^{(4)}$ curvature  of equation \eqref{effe} has no $\psi$ terms in its on-shell parametrization so that its space-time field-strength is simply $F^{(4)}_{\mu_1,\dots \mu_4}$. This implies that the field-strength of the 3-form $A^{(3)}$}, taking into account the definition \eqref{supcur} of the supercurvature $F^{(4)}$, is
\begin{align}
D_{[\mu_1} A^{(3)}_{\mu_2\mu_3\mu_4]}= F^{(4)}_{\mu_1,\dots \mu_4}+\frac12\bar\psi_{[\mu_1}\Gamma_{\mu_2\mu_3}\psi_{\mu_4]}.
\end{align}
Finally, using the parametrization of the curvature $R^{ab}$ we have
\begin{align}
& R^{ab}_{~~\mu\nu}=R^{ab}_{~~mn }V^m_\mu\,V^n_\nu +\Theta^{ab}_{[\mu}\psi_{\nu]}+\psi_{[\mu}\Gamma_{mn}\psi_{\nu]} F^{mnab}+\frac{1}{4!}
\psi_{[\mu}\Gamma^{abpqrs} \psi_{\nu]} F_{pqrs},
\end{align}
and contracting indices in equation \eqref{ra} one obtains the Einstein equation.

Using these tools one can derive the space-time Lagrangian and the equations of motion which, modulo conventions, coincide with those given in the celebrated Cremmer--Julia--Scherk $d=11$ supersymmetric space-time approach \cite{Cremmer:1978km}. \footnote{Note that also in \cite{Cremmer:1978km} supercovariant curvatures are introduced. }

\sect{Conclusions}

In this paper we have focused on the duality between Free Differential Algebras and $L_\infty$ algebras, as exemplified in supergravity theories containing higher forms. The case of $d=11$ supergravity has been discussed in detail, and its $L_3$ and $L_6$ structure explicitly
presented.

Comprehensive discussions on the applications of the group manifold method
for the construction of supergravity theories in diverse dimensions can be found in the recent reviews \cite{gm24,gm25,handbook}.
\sk
\noi We list here some of the advantages/motivations:
\sk
\noi - all fields have a group-geometric origin, even if not all of them are
 gauge fields.

 \noi - all symmetries have a common origin as diffeomorphisms on $\Gtilde$.

\noi  - there is a systematic procedure based on group geometry to construct
  actions, invariant under diffeomorphisms, and under gauge symmetries closing
  on a subgroup of $G$.

\noi   - supersymmetry is formulated in a very natural way as a
  diffeomorphism in Grassmann directions of a supermanifold.

\noi - closer contact is maintained with the usual component actions, whereas in the superfield formalism the actions look quite different.
  In fact the group manifold action interpolates between the component and the superfield actions of the same supergravity theory, see \cite{if1,if2,if3,if4}.

\noi - in the group manifold formulation of $d=6$ supergravity \cite{d6SG} and $d=10$ supergravity
\cite{d10SG} the selfdual conditions for the 3-form (in $d=6$) and 5-form (in $d=10$) curvatures are a yield of the field equations
in the respective superspaces, and do not need to be imposed as external constraints.
\sk
\noi Finally, we recall some conceptual advances due to the group-geometric treatment of supergravity:
\sk
\noi - the generalization to $p$-form
potentials, necessary to treat supergravity theories with $p$-form fields, in the framework of
Free Differential Algebras (FDA) \cite{sullivan,DFd11,CDF,DFTvN,FDAqwm}, in the dual framework of $L_\infty$ algebras discussed in the present paper,
and in the alternative dual formulation of \cite{FDAdual1,FDAdual2,FDAdual3,FDAdual4}.

\noi - the bridge between superspace and group manifold methods provided by
superintegration, developed in ref.s \cite{if1,if2,if3,if4}.

\noi - a covariant hamiltonian formalism, initially proposed in \cite{CCF1,CCF2,CCF3}, based on the definition of field momenta as
derivatives of the Lagrangian with respect to the exterior derivative of the fields, not involving
a preferred direction (time). Recent developments \cite{CD,SGcovariantH} include the construction
of all canonical symmetry generators for $N=1$, $d=4$ supergravity \cite{SGcovariantH}. This
covariant hamiltonian formalism can also be generalized to a noncommutative (twisted)
setting \cite{LCtwistedH}, describing noncommutative twisted (super)gravity \cite{AC1,AC2}.

\section*{Acknowledgements}

We acknowledge partial support from INFN (CSN4, Iniziativa Specifica GSS).
This research has a financial support from Universit\`a del Piemonte Orientale.

\appendix

\sect{Super-Poincar\'e Lie Algebra in $d=4$}

The $d=4$ super-Poincar\'e Lie algebra is defined by the following commutation relations
between the translations $P_a$, Lorentz rotations $M_{ab}$ and supersymmetry
generators $\Qbar_\alpha$:
 \eqa
 & &  [P_a,P_b] =0  \label{PoincarePP} \\
 & & [M_{ab},M_{cd}]= -{1 \over 2} (\eta_{ad} M_{bc} + \eta_{bc} M_{ad} -\eta_{ac} M_{bd} -\eta_{bd} M_{ac}) \label{PoincareMM} \\
 & & [M_{ab},P_c]= -{1 \over 2} ( \eta_{bc} P_{a} - \eta_{ac} P_{b}) \label{PoincareMP} \\
 & & [P_a,\Qbar_\al]=0 \\
 & & [M_{ab},\Qbar_\beta]= -{1 \over 4} \Qbar_\alpha (\gamma_{ab})^\alpha_{~\beta}   \label{sPoincareMQ} \\
 & & \{ \Qbar_\alpha,\Qbar_\beta \} = -i (C\gamma^a)_{\alpha \beta}  P_{a}, \label{sPoincareQQ}
  \ena
where $\eta$ is the flat Minkowski metric, and $C_{\al\be} $ the charge conjugation matrix.  The spinorial generator $\Qbar_\alpha  \equiv Q^\beta
C_{\beta\alpha}$ is a Majorana spinor, i.e. $Q^\beta C_{\beta\alpha} = Q^\dagger_\beta (\gamma_0)^\beta_{~\alpha}$.

\vfill\eject
\end{document}